\documentclass[aps,prl,twocolumn,superscriptaddress,floatfix,longbibliography]{revtex4-2}
\usepackage{graphicx}% Include figure files
\usepackage{dcolumn}% Align table columns on decimal point
\usepackage{bm}% bold math
\usepackage{bbold}
\usepackage{hyperref}% add hypertext capabilities
\usepackage{amsmath}
\makeatletter
\begin{document}

%\preprint{APS/123-QED}

\title{Fundamental Sensitivity Limits for non-Hermitian Quantum Sensors}


\author{Wenkui Ding}
\affiliation{Beijing National Laboratory for Condensed Matter Physics, Institute of
Physics, Chinese Academy of Sciences, Beijing 100190, China}
\affiliation{Department of Physics, Zhejiang Sci-Tech University, 310018 Zhejiang, China}
\author{Xiaoguang Wang}
\email{xgwang@zstu.edu.cn}
\affiliation{Department of Physics, Zhejiang Sci-Tech University, 310018 Zhejiang, China}
\author{Shu Chen}
\email{schen@iphy.ac.cn}
\affiliation{Beijing National Laboratory for Condensed Matter Physics, Institute of
Physics, Chinese Academy of Sciences, Beijing 100190, China}
\affiliation{School of Physical Sciences, University of Chinese Academy of Sciences,
Beijing, 100049, China}
\begin{abstract}
Considering non-Hermitian systems implemented by utilizing enlarged quantum systems, we determine the fundamental limits for the sensitivity of non-Hermitian sensors from the perspective of quantum information.
We prove that non-Hermitian sensors do not outperform their Hermitian counterparts (directly couple to the parameter) in the performance of sensitivity, due to the invariance of the quantum information about the parameter.
By scrutinizing two concrete non-Hermitian sensing proposals, which are implemented using full quantum systems, we demonstrate that the sensitivity of these sensors is in agreement with our predictions. Our theory offers a comprehensive and model-independent framework
for understanding the fundamental limits of non-Hermitian quantum sensors and builds the bridge over the gap between non-Hermitian physics and quantum metrology.
\end{abstract}

\date{\today}% It is always \today, today,
\maketitle

%\tableofcontents

%\section{Introduction}
\textit{Introduction.--} Parallel with the rapid development in quantum technology, quantum metrology~\cite{giovannetti2006quantum,pezze2018quantum,braun2018quantum,rams2018at} and quantum sensing~\cite{degen2017quantum,barry2020sensitivity} are becoming one of the focuses in quantum science.
Quantum sensors exploit quantum coherence or quantum correlations to detect weak or nanoscale signals and exhibit great advantages in accuracy, repeatability and precision.
%Quantum sensors make use of quantum coherence and correlations to detect weak or nanoscale signals, offering significant advantages in terms of accuracy, repeatability, and precision.
Recently, a number of sensing proposals utilizing novel properties of non-Hermitian physics~\cite{bergholtz2021exceptional,el2018non,quantum2022luo} have been proposed and experimentally demonstrated.
%In recent years, several proposals for sensing utilizing the unique properties of non-Hermitian physics have emerged and been experimentally demonstrated~\cite{bergholtz2021exceptional,el2018non,kononchuk2022exceptional}.
For example, non-Hermitian lattice systems with skin effect~\cite{budich2020non,koch2022quantum} or non-reciprocity~\cite{mcdonald2020exponentially} have been suggested to realize enhanced sensing.
Specifically, the divergence of the susceptibility near the exceptional point (EP) is exploited to realize enhanced sensing with arbitrary precision~\cite{wiersig2014enhancing,wiersig2016sensors,ren2017ultrasensitive,sunada2017large} and it has been demonstrated using various classical (quasi-classical) physical systems~\cite{liu2016metrology,chen2017exceptional,hodaei2017enhanced,lai2019observation,kononchuk2022exceptional} or quantum systems~\cite{yu2020experimental,wang2020petermann}.
%Specifically, the divergence of the susceptibility near the exceptional point (EP) is exploited to realize enhanced sensing with arbitrary precision~\cite{wiersig2014enhancing,wiersig2016sensors,ren2017ultrasensitive,sunada2017large} and it has been demonstrated using various classical~\cite{liu2016metrology,chen2017exceptional,hodaei2017enhanced,lai2019observation} or quantum systems~\cite{yu2020experimental,wang2020petermann}.
While these early experiments claimed enhancements compared to conventional Hermitian sensors, subsequent theoretical work has cast doubt on these results \cite{langbein2018no,lau2018fundamental,zhang2019quantum,chen2019sensitivity,duggan2022limitations},
suggesting that the reported enhancements may not have fully taken into account the effects
of noise.
%Actually, the sensitivity or precision is defined in terms of signal-to-noise ratio.
%When the noise is taken into account, which is essential for the standard definition of sensitivity for sensors, this enhancement may disappear.
After taking into account the noise, some theoretical works show the enhancement in
sensitivity provided by non-Hermitian sensors may disappear~\cite{langbein2018no,chen2019sensitivity}.
However, other theoretical works have claimed that the enhancement can persist even in the presence
of noise~\cite{lau2018fundamental,zhang2019quantum}.
%Very recently, there are experiments on the non-Hermitian sensing schemes that take into account the effect of noise, some of them show enhancement in the sensitivity~
While some recent experiments have demonstrated enhanced sensitivity despite
the presence of noise
\cite{yu2020experimental,kononchuk2022exceptional}, others have shown no
such enhancement~\cite{wang2020petermann}.
%Meanwhile, recent experiments on non-Hermitian sensing schemes have investigated the impact of noise on sensitivity enhancement. While some experiments have demonstrated enhanced sensitivity despite the presence of noise~\cite{yu2020experimental}, others have shown no such enhancement~\cite{wang2020petermann}.
Currently, the fundamental limitations
imposed by noise on non-Hermitian sensors are
still a topic of debate~\cite{review2020jan}, and a definitive conclusion on whether the non-Hermitian physics is superior for sensing is still elusive.

In sensing schemes that rely on quantum systems,
quantum noise always arises during the projective measurement
of the parameter-dependent quantum state~\cite{itano1993quantum}.
This noise originates from quantum mechanics and cannot be eliminated, leading to the fundamental sensitivity limit.
Quantum metrology focuses on how to beat the standard quantum limit by employing quantum correlations, like entanglement or squeezing~\cite{pezze2018quantum}.
%Usually, non-Hermitian physics is used to approximate open system dynamics, 
While non-Hermitian systems can serve as
an effective description of open system dynamics in certain
situations~\cite{breuer2002theory,el2018non}, the decoherence and dissipation in open systems are detrimental to the useful quantum features required for metrology 
~\cite{escher2011general,demkowicz2012elusive,chin2012quantum,alipour2014quantum,beau2017nonlinear}.
Therefore, the sensitivity enhancement from non-Hermitian sensors, which can be embedded in open systems, is quite counter-intuitive.
%Actually, there have been argues that EP sensors cannot make sensing enhancements, since the coalescence of the eigenstates compensates the divergent susceptibility of the eigenenergy~\cite{langbein2018no,lau2018fundamental,zhang2019quantum,chen2019sensitivity,duggan2022limitations}.
Various theoretical works have been devoted to analyze the effect from the noise ~\cite{langbein2018no,lau2018fundamental,zhang2019quantum,chen2019sensitivity,duggan2022limitations}, however, these investigations usually require modeling the effect of noise and calculating the dynamics using tools such as the quantum Langevin equation, for specific sensing schemes and probe states.
Here, we provide a general conclusion on the fundamental sensitivity limit from the perspective of quantum information~\cite{lu2010quantum}, without the requirement to solve intricate non-unitary quantum dynamics and independent of specific noise forms, probe states, and measurement regimes. 
We unambiguously prove that the non-Hermitian quantum sensors do not surpass the ultimate sensitivity of their Hermitian counterparts and cannot achieve arbitrary precision in realistic experimental settings with finite quantum resources.

%\section{Sensitivity bound for a general parameter encoding Hamiltonian}
\textit{Sensitivity bound for unitary parameter encoding.--}Quantum metrology or quantum parameter estimation is to estimate the parameter $\lambda$ from the parameter-dependent quantum state $\rho_\lambda$. One crucial step is to make measurements on the quantum state.
The measurement can be described by a Hermitian operator $\Pi$, and the probability of obtaining the measurement outcome $\xi$, conditioned on the parameter $\lambda$, is $P(\xi|\lambda)=\text{Tr}(\Pi\rho_\lambda)$.
We can evaluate the classical Fisher information $\mathcal{I}_\lambda$ corresponding to this specific measurement as $\mathcal{I}_\lambda=\sum_{\xi} P(\xi|\lambda)\left(\frac{\partial\ln P(\xi|\lambda)}{\partial \lambda}\right)^2$,which reflects the amount of information about the parameter contained in the distribution of measurement outcomes.
%\begin{equation}
%\mathcal{I}_\lambda=\sum_{\xi} P(\xi|\lambda)\left(\frac{\partial\ln P(\xi|\lambda)}{\partial \lambda}\right)^2,
%\end{equation}
Meanwhile, the estimation uncertainty is given by $\delta^2\lambda=\left\langle \left( \frac{\lambda_{\text{est}}}{d\langle\lambda_{\text{est}}\rangle/d\lambda}-\lambda\right)^2\right\rangle$,
where $\lambda_{\text{est}}$ is the estimated value when the number of probes ($N$) and the number of trials ($\nu$) are finite, while $\lambda$ is the true value of the parameter.
For the unbiased estimator, we have $d\langle\lambda_{\text{est}}\rangle/d\lambda=1$.
In fact, the classical Fisher information bounds the estimation uncertainty achievable in this specific measurement, which fulfills the so-called Cram\'{e}r-Rao bound: $\delta \lambda\geq {1}/{\sqrt{\nu \mathcal{I}_\lambda}}$, where $\nu$ is the number of repetitions or trials.
This bound can be attained asymptotically as $\nu\rightarrow\infty$.
When it is optimized over all possible measurements, we can find the maximal value of the classical Fisher information, known as the quantum Fisher information (QFI)~\cite{braunstein1994statistical}, $\mathcal{I}_\lambda\leq F_\lambda$.
Accordingly, the ultimate precision of parameter estimation for a specific parameter-dependent quantum state can be determined using the quantum Cram\'{e}r-Rao bound~\cite{helstrom1967minimum}, $\delta\lambda\geq{1}/{\sqrt{\nu F_\lambda}}$.
The QFI~\footnote{Equivalently, the quantum Fisher information can be calculated as $F_\lambda=-4\left.\frac{\partial^2\mathcal{F}_Q(\rho_\lambda,\rho_{\lambda+\delta_\lambda})}{\partial\delta_\lambda^2}\right|_{\delta_\lambda=0}$, where the quantum fidelity between quantum states is defined as $\mathcal{F}_Q(\rho_\lambda,\rho_{\lambda+\delta_\lambda})\equiv\text{Tr}\left[\sqrt{\sqrt{\rho_{\lambda+\delta_\lambda}}\rho_\lambda\sqrt{\rho_{\lambda+\delta_\lambda}}}\right]$.} can be determined as $F_\lambda=\text{Tr}[\rho_\lambda\mathcal{L}^2]$, where $\mathcal{L}$ is the symmetric logarithmic derivative defined by ${\partial\rho_\lambda}/{\partial \lambda}=(\mathcal{L}\rho_\lambda+\rho_\lambda\mathcal{L})/2$.

Usually, the parameter-dependent quantum state $\rho_\lambda$ is obtained through time evolution governed by the parameter-dependent Hamiltonian $\hat{H}_\lambda(t)$.
To be more specific, with the parameter-independent initial state (probe state) $\rho_0$, the parameter encoding process can be described as $\rho_\lambda(t)=U_\lambda(0\rightarrow t)\rho_0U^\dagger_\lambda(0\rightarrow t)$, where the unitary time evolution operator $U_\lambda(0\rightarrow t)=\mathcal{T}e^{-i\int_{0}^t{\hat{H}_\lambda(s)ds}}$, with $\mathcal{T}$ being the time-ordering operator.
In the case where the initial state is a pure state, $\rho(0)=|\Psi_0\rangle\langle\Psi_0|$, the QFI can be calculated as $F_\lambda(t)=4(\langle \Psi_0|h_\lambda^2(t)|\Psi_0\rangle-\langle\Psi_0|h_\lambda(t)|\Psi_0\rangle^2)\equiv 4\text{Var}[h_\lambda(t)]|_{|\Psi_0\rangle}$, where the Hermitian operator $h_\lambda(t)\equiv iU_\lambda^\dagger(0\rightarrow t)\frac{\partial}{\partial \lambda}U_\lambda(0\rightarrow t)$ is called the transformed local generator~\cite{pang2014quantum,liu2015quantum}.
We have defined $\text{Var}[{\hat{A}}]|_{|\Psi\rangle}$ as the variance of the Hermitian operator ${\hat{A}}$ with respect to $|\Psi\rangle$.
It satisfies $\text{Var}[\hat{A}]|_{|\Psi\rangle}\leq ||\hat{A}||^2/4$ for arbitrary $|\Psi\rangle$~\cite{boixo2007generalized},  where the seminorm is defined as $||\hat{A}||\equiv M_A-m_A$, with $M_A$ ($m_A$) being the maximum (minimum) eigenvalue of $\hat{A}$.
Then it follows $F_\lambda(t) \leq ||h_\lambda(t)||^2\equiv F_\lambda^{(c)}(t)$, where $F_\lambda^{(c)}(t)$ is defined as the channel QFI, corresponding to the maximum QFI achievable by optimizing over all possible probe states.

The triangle inequality for the seminorm of Hermitian operators~\cite{boixo2007generalized} states that $||\hat{A}+\hat{B}||\leq ||\hat{A}||+||\hat{B}||$.
Using the definition of $h_\lambda$ and the Schr\"{o}dinger equation $i\partial U_\lambda/\partial t=H_\lambda U_\lambda$, we can obtain $\frac{\partial h_\lambda}{\partial t}=U_\lambda^\dagger(0\rightarrow t)\frac{\partial H_\lambda(t)}{\partial\lambda}U_\lambda(0\rightarrow t)$.
Thus, the transformed local generator can be explicitly represented as $h_\lambda(t)=\int_0^t{U_\lambda^\dagger(0\rightarrow s)\frac{\partial H_\lambda(s)}{\partial \lambda}U_\lambda(0\rightarrow s)ds}$.
By applying the triangle inequality, we obtain $||h_\lambda(t)||\leq\int_0^t{\left|\left|U_\lambda^\dagger(0\rightarrow s)\frac{\partial H_\lambda(s)}{\partial \lambda}U_\lambda(0\rightarrow s)\right|\right|ds}=\int_0^t{\left|\left|\frac{\partial H_\lambda(s)}{\partial \lambda}\right|\right|ds}$, where we have used the fact that unitary transformations do not change the spectrum of an operator.
Therefore, the upper bound of the channel QFI can be obtained as follows~\footnote{In particular, when $H_1$ is time-independent and the parameter is a multiplicative factor, i.e., $H_\lambda(t)=\lambda H_1+H_0(t)$, we reproduce the result in Ref.~\cite{boixo2007generalized} that, $F_\lambda\leq ||H_1||^2t^2$.
Specifically, for the well-researched case where $H_1=\sum_{j=1}^N h_j$, the operator seminorm $||H_1||=N(\varepsilon_M-\varepsilon_m)$, where $\varepsilon_M$ ($\varepsilon_m$) represents the maximum (minimum) eigenvalue of the one-body Hamiltonian $h_j$.
Consequently, the ultimate quantum Fisher information $\propto N^2$, corresponding to the well-known Heisenberg scaling.
Additionally, if $H_1$ contains $k$-body interaction terms, the seminorm may scale as $||H_1||\propto N^{k}$, leading to super-Heisenberg scaling~\cite{boixo2007generalized,rams2018at}.}:
\begin{equation}
\begin{aligned}
F_\lambda^{(c)}(t) \leq \left[\int_0^t\left|\left|\frac{\partial H_\lambda(s)}{\partial \lambda}\right|\right|ds\right]^2.
\end{aligned}
\end{equation}
Due to the convexity of QFI, the optimal probe state is always a pure state~\cite{fiderer2019maximal}.
Therefore, this bound is naturally applicable for mixed probe states.
Similar relations~\cite{pang2017optimal,fiderer2019maximal} have been obtained using different methods and have been employed to discuss unitary parameter encoding processes governed by Hermitian time-dependent Hamiltonians.
Furthermore, by utilizing the quantum Cram\'{e}r-Rao bound, we obtain the lower bound for the estimation uncertainty as follows~\footnote{In quantum sensing, the sensitivity is defined as the minimal detectable signal $\lambda_{\text{min}}$ that yields a unit signal-to-noise ratio for a unit integration time (sensing time)~\cite{degen2017quantum}.
Therefore, the sensitivity $\lambda_\text{min}$ is bounded by same inequality in Eq.~(\ref{eq:estimation_bound}), with the substitution $t=1\text{ s}$}:
\begin{equation}
\label{eq:estimation_bound}
\delta\lambda\ge \frac{1}{\sqrt{\nu}\int_0^t\left|\left|\frac{\partial H_\lambda(s)}{\partial \lambda}\right|\right|ds}.
\end{equation}
Here, we realize that this relation is actually not limited to unitary parameter encoding processes. Instead, this bound can be applied to investigate non-unitary parameter encoding processes, particularly in the context of open quantum systems or dynamics governed by non-Hermitian Hamiltonians.
%This relation is rather universal and can be used to determine the lower sensitivity bound for various types of quantum sensors, regardless of whether they are based on unitary or non-unitary parameter encoding processes.

In this Letter, we proceed further to investigate the bound on the change rate of the QFI.
By the definition of QFI, we obtain that $\frac{\partial F_\lambda}{\partial t}=8\left.\text{Cov}[\frac{\partial h_\lambda}{\partial t},h_\lambda]\right|_{|\Psi_0\rangle}$, where the covariance is defined as $\left.\text{Cov}[\hat{A},\hat{B}]\right|_{|\Psi\rangle}\equiv\frac{1}{2}\langle \Psi|\hat{A} \hat{B}+\hat{B} \hat{A} |\Psi\rangle-\langle \Psi|\hat{A} |\Psi\rangle\langle \Psi|\hat{B} |\Psi\rangle$.
The covariance inequality deduced from the Cauchy-Schwarz inequality states that $\left|\text{Cov}[\hat{A},\hat{B}]\right|\leq \sqrt{\text{Var}(\hat{A})\text{Var}(\hat{B})}$.
Applying this inequality, we find:
\begin{equation}
\begin{aligned}
\left|\text{Cov}[\frac{\partial h_\lambda}{\partial t},h_\lambda(t)]|_{|\Psi_0\rangle}\right|&\leq \sqrt{\text{Var}[U_\lambda^\dagger \frac{\partial H_\lambda}{\partial\lambda} U_\lambda]|_{|\Psi_0\rangle}}\frac{F_\lambda^{1/2}(t)}{2}\\
&\leq\frac{||\frac{\partial H_\lambda}{\partial\lambda}||}{2}\frac{F_\lambda^{1/2}(t)}{2}.
\end{aligned}
\end{equation}
After some algebra~\footnote{See Supplemental Material at [url] for the derivation of the bound of the change rate of the quantum Fisher information, the rate of dynamic quantum Fisher information for the pseudo-Hermitian quantum sensor and the derivation of the population fluctuation for the two-level system, which includes Ref. [30] and Ref. [43].}, we prove the following inequality:
\begin{equation}
\label{eq:qfi_rate}
\left|\frac{\partial F_\lambda^{1/2}(t)}{\partial t}\right|\leq \left|\left|\frac{\partial H_\lambda(t)}{\partial \lambda}\right|\right|.
\end{equation}
Namely, the change rate of the square root of QFI is only bounded by the spectral width of the derivative of the Hamiltonian with respect to the parameter.
$|{\partial F_\lambda^{1/2}(t)}/{\partial t}|$ measures how fast the quantum information about the parameter flows into or out of the quantum state.
It indicates that the quantum parameter encoding process cannot be accelerated by adding auxiliary parameter-independent Hamiltonian extensions.

%\section{open system and non-Hermitian quantum sensing}
\textit{Open system and non-Hermitian quantum sensing.--}In many situations, such as dynamics in open quantum systems or systems governed by non-Hermitian Hamiltonians, the dynamical process used to encode the parameter may be non-unitary.
%However, non-unitary dynamical process in open systems can generally be mapped to unitary dynamics in an enlarged Hilbert space by introducing extra degrees of freedom corresponding to the environment.
However, it is often possible to map these non-unitary processes to equivalent unitary dynamics in an enlarged Hilbert space, by introducing extra degrees of freedom that correspond to the environment~\cite{escher2011general}.
We now make this statement more rigorous for non-unitary sensing schemes.
Prior to applying the perturbation that incorporates the parameter to be estimated, the dynamical process in the open quantum system or non-Hermitian system, $R_S: \rho_S(0)\rightarrow \rho_S(t)$, can be mapped from a unitary evolution in an enlarged system, $\mathcal{M}(U_{S,E})\rightarrow R_S$.
This unitary time evolution operator for the combined system corresponds to a Hermitian Hamiltonian, $U_{S,E}\rightarrow \tilde{H}_\text{tot}$.
This Hamiltonian, $\tilde{H}_\text{tot}=H_S(t)+H_E(t)+H_{SE}(t)$, generally contains terms that describe the system $H_S(t)$, the environment $H_E(t)$ and the system-environment interaction $H_{SE}(t)$.
Subsequently, we introduce the perturbation that incorporates the parameter dependence.
In most scenarios, including the examples discussed in this work and various non-Hermitian sensing protocols, the parameter of interest directly couples to the degrees of freedom of the system and the perturbation can be represented by a Hermitian Hamiltonian $H_1(\lambda,t)$.
As a result, the overall parameter encoding process, corresponding to the dynamical evolution in the open system or non-Hermitian system, can be mapped to a unitary dynamics governed by a Hermitian Hamiltonian $H_\lambda(t)=\tilde{H}_\text{tot}+H_1(\lambda,t)$.
By mapping the dynamics to an enlarged system, we circumvent the analysis of intricate non-unitary parameter encoding processes.
By resorting to the corresponding unitary evolution in the enlarged system, we can straightforwardly apply the ultimate sensitivity bound in Eq.~(\ref{eq:estimation_bound}) and the QFI rate bound in Eq.~(\ref{eq:qfi_rate}).

Since the estimation parameter only associates with the degree of freedom of the system, we have $\partial H_\lambda/\partial\lambda=\partial H_1/\partial\lambda$.
Thus, the bounds in Eq.~(\ref{eq:estimation_bound}) and~(\ref{eq:qfi_rate}) reveal an intriguing insight: the ultimate sensitivity cannot be improved by coupling the system to the environment or by introducing auxiliary Hamiltonians.
%since the amount of information about the estimation parameter or the rate of information encoding is not increased.
This is because these additional factors do not increase the amount of information about the parameter or the rate of information encoding.
Correspondingly, the non-Hermitian sensor will not outperform its Hermitian counterpart in terms of ultimate sensitivity.
We now substantiate this conclusion by analyzing some concrete examples.

\begin{figure}
\includegraphics[width=0.5\textwidth]{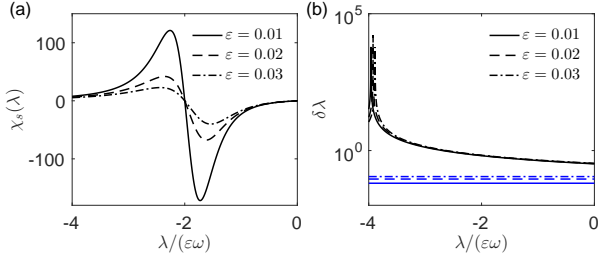}
\centering
\caption{\label{fig:example1} (a) Susceptibility of the normalized population with respect to $\lambda$ for different values of $\varepsilon$. It indicates that the maximal susceptibility diverges as $\varepsilon$ approaches zero. (b) Sensitivity corresponding to the measurement of the population in the state $|0\rangle_a\otimes|0\rangle_s$. It indicates that the sensitivity at the optimal measurement point (corresponding to the maximal susceptibility) does not diverge when $\varepsilon$ approaches zero. The blue lines represent the sensitivity bound of the Hermitian counterpart.}
\end{figure}

%\section{example I: single-qubit pseudo-Hermitian sensor}
\textit{Example I: single-qubit pseudo-Hermitian sensor.--}A single-qubit pesudo-Hermitian~\footnote{A non-Hermitian Hamiltonian $H$ is said to be pseudo-Hermitian if $\eta H=H^\dagger \eta$, where $\eta$ is a Hermitian invertible linear operator.} Hamiltonian, described by
\begin{equation}
\hat{H}_s=\mathcal{E}_\lambda\left(\begin{array}{ccc}
0&\delta_\lambda^{-1}\\
\delta_\lambda&0
\end{array}\right),
\end{equation}
is employed to realize enhanced quantum sensing in Ref.~\cite{chu2020quantum}, where $\mathcal{E}_\lambda$ and $\delta_\lambda$ depend on the parameter $\lambda$ that is being estimated.
According to the Naimark dilation theory~\cite{naimark2008gunther,kawabata2017information}, a dilated two-qubit system with a properly prepared initial state can be used to simulate the dynamics of this pseudo-Hermitian Hamiltonian, conditioned on the post-selection measurement of the ancilla qubit~\cite{wu2019observation}.
The Hermitian Hamiltonian of this dilated two-qubit system is
\begin{equation}
\hat{H}_\text{tot}=b\hat{I}^{(a)}\otimes \hat{\sigma}_x^{(s)}-c\hat{\sigma}_y^{(a)}\otimes\hat{\sigma}_y^{(s)}+\lambda\hat{I}^{(a)}\otimes\hat{\sigma}_x^{(s)},
\end{equation}
where $\hat{\sigma}_{\alpha=x,y,z}^{(s)}$ ($\hat{\sigma}_{\alpha=x,y,z}^{(a)}$) represents the Pauli operators of the system qubit (ancilla qubit).
The coefficients $b={4\omega\varepsilon(1+\varepsilon)}/{(1+2\varepsilon)}$ and $c={2\omega\sqrt{\varepsilon(1+\varepsilon)}}/{(1+2\varepsilon)}$, where $\varepsilon$ and $\omega$ describe the qubit.
This specific dilated Hamiltonian can be mapped to $\hat{H}_s$, with $\mathcal{E}_\lambda=\sqrt{(b+\lambda)^2+c^2}$ and $\delta_\lambda={(\lambda+2\varepsilon\omega)}/{\mathcal{E}_\lambda}$.
The time evolution of the quantum state governed by $\hat{H}_s$ is $|\psi\rangle_s=e^{-i\hat{H}_st}|0\rangle_s=
\cos(\mathcal{E}_\lambda t)|0\rangle_s-i\delta_\lambda\sin(\mathcal{E}_\lambda t)|1\rangle_s$.
Thus the normalized population in $|0\rangle_s$ is $S(\lambda,t)={1}/{[1+\delta_\lambda^2\tan^2{(\mathcal{E}_\lambda t)}]}$.
In Fig.~\ref{fig:example1}(a), we plot the susceptibility $\chi_s(\lambda)\equiv \partial S/\partial \lambda$ as a function of $\lambda$ for a fixed evolution time $t=\tau\equiv{\pi}/{[4\omega\sqrt{\varepsilon(1+\varepsilon)}]}$.
The result indicates that the maximal value of the susceptibility diverges as $\varepsilon\rightarrow 0$, which corresponds to the eigenstate coalescence.
Based on this feature, the authors in Ref.~\cite{chu2020quantum} proposed the pseudo-Hermitian enhanced quantum sensing scheme.

On the other hand, for the dilated two-qubit system, the probe state should be prepared as $|\Psi_0\rangle=\left(\sqrt{\frac{1+\varepsilon}{1+2\varepsilon}}|0\rangle_a+\sqrt{\frac{\varepsilon}{1+2\varepsilon}}|1\rangle_a\right)\otimes|0\rangle_s$ in order to correctly simulate the non-Hermitian dynamics.
The normalized population $S(\lambda,t)$ actually corresponds to the probability that the system qubit is in state $|0\rangle_s$, conditioned on the ancilla qubit being in state $|0\rangle_a$.
Equivalently, by calculating the dynamics of the total system $|\Psi(\tau)\rangle=e^{-i\hat{H}_{\text{tot}}\tau}|\Psi_0\rangle$, we can directly evaluate the probability in state $|0\rangle_a\otimes|0\rangle_s$ as
\begin{equation}
\begin{aligned}
P_1=\frac{1+\varepsilon}{1+2\varepsilon}\cos^2\left[t\sqrt{\lambda^2+\frac{8\varepsilon(1+\varepsilon)\lambda\omega}{1+2\varepsilon}+4\varepsilon(1+\varepsilon)\omega^2}\right].
\end{aligned}
\end{equation}
Due to the quantum projection noise, there is uncertainty in the determination of $P_1$.
This uncertainty originates from the quantum projective measurement and follows a binomial distribution.
The variance of the estimated probability is $\text{Var}[\hat{P}_1]={P_1(1-P_1)}/{\nu}$, where $\nu$ is the number of trials (repetitions)~\cite{Note4}.
Using the error propagation formula, we can evaluate the estimation uncertainty for this specific sensing scheme as $\delta\lambda={\sqrt{\text{Var}[\hat{P}_1]}}/{|\frac{\partial P_1}{\partial \lambda}|}$.
We plot the sensitivity in Fig.~\ref{fig:example1}(b), which shows no divergence at the corresponding divergent positions of $\chi_s(\lambda)$ in Fig.~\ref{fig:example1}(a).
This absence of divergence in the sensitivity is attributed to the fact that the divergence in $\chi_s(\lambda)$ when $\varepsilon\rightarrow 0$ is accompanied by a vanishing success probability in the post-selection measurement.
Namely, most experimental trails fail to provide useful information on the parameter.
As a comparison, the counterpart Hermitian sensor simply employs $\hat{V}=\lambda\hat{I}^{(a)}\otimes\hat{\sigma}_x^{(s)}$ as the parameter encoding generator. 
The sensitivity bound in Eq.~(\ref{eq:estimation_bound}) indicates $\delta\lambda\geq\frac{1}{\sqrt{\nu}\tau||\hat{\sigma}_x||}=\frac{1}{2\sqrt{\nu}\tau}$.
We plot this ultimate sensitivity bound in Fig.~\ref{fig:example1}(b) as the blue lines, indicating that the non-Hermitian sensor does not outperform its Hermitian counterpart.
Furthermore, the rate of dynamic QFI can be calculated exactly~\cite{Note4} as follows:
\begin{equation}
\frac{\partial F^{1/2}_\lambda(t)}{\partial t}=2\frac{\cos^2\theta+\sin^2\theta\frac{\sin{(2\Omega t)}}{2\Omega t}}{\sqrt{\cos^2\theta+\sin^2\theta\frac{\sin^2(\Omega t)}{(\Omega t)^2}}},
\end{equation}
where we define $(b+\lambda)/\Omega=\cos\theta$ and $c/\Omega=\sin\theta$.
It follows that $-2\leq\frac{\partial F^{1/2}_\lambda(t)}{\partial t}\leq 2$, which verifies our theory in Eq.~(\ref{eq:qfi_rate}).

\begin{figure}
\includegraphics[width=0.5\textwidth]{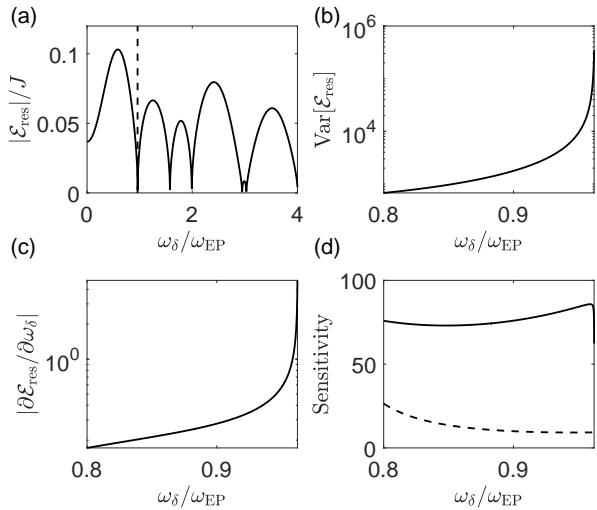}
\centering
\caption{\label{fig:example2} (a) Response energy shows sharp dips near the EP for the periodically driven non-Hermitian system. The dashed line indicates the position of the first EP. In (b-d), the range of $\omega$ corresponds to the zoom-ins of the left side of the first EP. (b) Variance of the response energy near the EP diverges. (c) Susceptibility of the response energy exhibit divergence near the EP. (d) The sensitivity, which is inversely proportional to the signal-to-noise ratio, shows no divergence. The dashed line represents the theoretical sensitivity bound of the Hermitian counterpart.}
\end{figure}

%\section{example II: EP based sensor using a single trapped ion}
\textit{Example II: EP based sensor using a single trapped ion.--}We now consider the sensor based on exceptional point realized in a dissipative single-qubit open system in Ref.~\cite{ding2021experimental}.
The sensing mechanism relies on an effective periodically driven~\cite{li2019observation} $\mathcal{PT}$-symmetric non-Hermitian Hamiltonian given by
\begin{equation}
\hat{H}_{\mathcal{PT}}=J[1+\cos(\omega t)]\hat{\sigma}_x+i\Gamma\hat{\sigma}_z,
\end{equation}
where $\hat{\sigma}_{x,z}$ are the Pauli operators, $J$ is the coupling strength, $\omega$ is the modulation frequency of the coupling strength, and $\Gamma$ is the dissipation rate.
Actually, the practically implemented Hamiltonian in the experiment is $\hat{H}_{\mathcal{PT}}^\prime=\hat{H}_{\mathcal{PT}}-i\Gamma\hat{I}$, which is a passive $\mathcal{PT}$-symmetric system with $\hat{I}$ being the identity operator.
The perturbation applied to the system is $\hat{H}_\delta=\frac{\delta}{2}\cos(\omega_\delta t)(\hat{I}-\hat{\sigma}_z)$, where $\delta$ and $\omega_\delta$ are the amplitude and frequency of the perturbation field, respectively, while $\omega_\delta$ is the parameter to be estimated.
After the system evolves from specific initial states for a duration of $T=2\pi/\omega$, we can determine the response energy $\mathcal{E}_{\text{res}}$ via $P_J(T)-P_\Gamma(T)=\sin^2(\mathcal{E}_{\text{res}}T)$.
Here, the measurable quantities are defined as $P_J(T)=\left|\left\langle \uparrow\right|U(T)\left|\downarrow\right\rangle\right|^2$ and $P_\Gamma(T)=\left|\frac{\left\langle\uparrow\right|-\left\langle\downarrow\right|}{\sqrt{2}}U(T)\frac{\left|\uparrow\right\rangle+\left|\downarrow\right\rangle}{\sqrt{2}}\right|^2$, with $U(T)=\mathcal{T}e^{-i\int_0^T{[\hat{H}_{\mathcal{PT}}(t)+\hat{H}_\delta(t)]dt}}$.
The absolute value of the response energy $\mathcal{E}_{\text{res}}$ as a function of $\omega_\delta$ is plotted in Fig.~\ref{fig:example2}(a)~\footnote{In fact, in order to realize the EP enhanced sensing, we should first determine the location of EP. To be more specific, initially, only $\hat{H}_{\mathcal{PT}}$ is applied (by setting $\hat{H}_\delta(t)=0$ temporarily) to determine the values of $\omega_{\text{EP}}$ and $\Gamma_{\text{EP}}$, which satisfy $P_J(T)-P_\Gamma(T)=0$, since the sign of $P_J-P_\Gamma$ serves as an indicator of the phase boundary~\cite{ding2021experimental}.}.
As it is shown, the response energy exhibits sharp dips near the EP~\footnote{Due to the same $\sqrt{\nu}$ scaling in the sensitivity of the non-Hermitian sensor and its Hermitian counterpart, we plot these figures for $\nu=1$.}. 
This characteristic feature has motivated the authors in Ref.~\cite{ding2021experimental} to suggest the sensing application, since a minor change in $\omega_\delta$ will result in a significant variation in the response energy.
Indeed, in Fig.~\ref{fig:example2}(c), we present the susceptibility $|\partial\mathcal{E}_{\text{res}}/\partial\omega_\delta |$ as a function of $\omega_\delta$ and it exhibits a divergence near the EP.

However, the study in Ref. [54] has neglected effects from the quantum noise.
Here, since $P_J$ and $P_\Gamma$ actually correspond to projective measurements on the spin state, the quantum projection noise will result in uncertainties in their determination.
The variance of the estimated $\hat{P}_J$ and $\hat{P}_\Gamma$ can be expressed as $\text{Var}[\hat{P}_i]={P_i(C_0-P_i)}/{\nu}, \text{with }i=J,\Gamma$, where $\nu$ is the number of trials and $C_0\equiv e^{2\Gamma T}$~\footnote{The constant $C_0$ is given by $C_0\equiv e^{2\Gamma T}$ because the evolution is governed by $H^\prime_{\mathcal{PT}}$ in the experiment.}.
To avoid the complication of dealing with complex response energies, we focus on the region near the EP where $P_J-P_\Gamma>0$.
Applying the theory of uncertainty propagation, we obtain the uncertainty in the estimation of the response energy as $\text{Var}[\hat{\mathcal{E}}_\text{res}]=\frac{1}{4\nu T^2}\frac{C_0(P_J+P_\Gamma)-(P_J^2+P_\Gamma^2)}{(P_J-P_\Gamma)(1-P_J+P_\Gamma)}$, where we have used the fact that measurements on $P_J$ and $P_\Gamma$ are independent.
We plot the variance of the measured response energy in Fig.~\ref{fig:example2}(b) as a function of $\omega_\delta$, and it shows that the uncertainty in the determination of $\mathcal{E}_\text{res}$ also diverges when $\omega_\delta$ approaches the EP.
The overall sensitivity can be evaluated as $\delta{\omega_\delta}={\sqrt{\text{Var}[\hat{\mathcal{E}}_{\text{res}}]}}/{|\frac{\partial\mathcal{E}_{\text{res}}}{\partial\omega_\delta}|}$, and we plot it in Fig.~\ref{fig:example2}(d).
It shows that the divergence of the susceptibility is completely compensated by the divergence of the uncertainty, resulting in an overall sensitivity without divergence when approaching the EP.
%On the other hand, according to our theory, the performance of the non-Hermitian sensor does not outperform its Hermitian counterpart.
On the other hand, the Hermitian counterpart simply uses $\hat{H}_\delta$ as the parameter encoding generator. According to Eq.~(\ref{eq:estimation_bound}), the ultimate sensitivity bound is given by $\delta{\omega_\delta}\geq \frac{\omega_\delta^4}{\sqrt{\nu}\delta^2[\sin(\omega_\delta T)-\omega_\delta T\cos(\omega_\delta T)]^2}$.
The dashed line in Fig.~\ref{fig:example2}(d) corresponds to this ultimate sensitivity bound. It also demonstrates that the ultimate precision of the Hermitian sensor always exceed the corresponding non-Hermitian sensor.

%\section{Discussion}
\textit{Summary and discussion.--}%In the non-Hermitian physics community,
In summary, we have unveiled the fundamental sensitivity limit for non-Hermitian sensors in the context of open quantum systems. Our results indicate clearly that non-Hermitian sensors do not outperform their Hermitian counterparts. 
In fact, when comparing the performance of quantum sensors, it is essential to fix the quantum resources consumed by these sensors. Actually, when resources are unlimited, even ideal Hermitian sensors can theoretically achieve arbitrary precision. However , in practical sensing scenarios, resources are always limited. The number of probes, sensing time, and the number of trials are examples of limited resources. As a result, achieving arbitrary precision is not possible in practical sensing scenarios.
The aforementioned instances are characterized by a single probe. Notably, although these cases exhibit divergence in certain measurable quantities, it does not imply that the sensitivity diverges, leading to `arbitrary precision', since the sensitivity of their Hermitian counterparts does not diverge (even without Heisenberg scaling for only $N=1$).

%In Ref.~\cite{yu2020experimental}, an EP enhanced sensing scheme is experimentally investigated using a quantum $\mathcal{PT}$-symmetric system assisted by weak measurement.
%The authors reported an enhancement of 8.856 times over the conventional Hermitian sensor.
%This enhancement is largely due to the sensing scheme for the Hermitian sensor is not optimal.
%Besides, non-Hermitian lattice systems are proposed to realize sensing schemes, utilizing the skin effect~\cite{budich2020non,koch2022quantum} or the non-reciprocity~\cite{mcdonald2020exponentially}, which claimed a sensitivity with an exponential scaling to the chain size.
%According to our theory, there should be no dependence on the chain size for the ultimate sensitivity, since it only depends on the dimension of the subsystem that directly couples to the parameter.

In Ref.~\cite{yu2020experimental}, a sensing scheme utilizing an experimentally realized $\mathcal{PT}$-symmetric system was reported to enhance the sensitivity by a factor of 8.856 over a conventional Hermitian sensor. However, this enhancement is probably attributed to the choice of non-optimal initial probe state used for the Hermitian sensor, and similarly these seeming sensitivity enhancements in Refs.~\cite{lau2018fundamental,zhang2019quantum} may not exist if making comparison over optimal probe states. Furthermore, non-Hermitian lattice systems utilizing the skin effect\cite{budich2020non,koch2022quantum} or the non-reciprocity~\cite{mcdonald2020exponentially}, have claimed exponential scaling of sensitivity with the lattice size.
%However, according to our theory, the ultimate sensitivity should not depend on the lattice size, as it is only determined by the dimension of the subsystem that directly couples to the parameter, although the sensitivity may still depend on the lattice size if the probe state or measurement is not optimal.
However, our theory shows that the ultimate sensitivity should not depend on the lattice size, as it is solely determined by the subsystem dimension that directly couples to the parameter. Nevertheless, for non-optimal probe states or measurements, the sensitivity may still depend on the lattice size.

%In a recent experimental study \cite{yu2020experimental}, an EP-enhanced sensing scheme was investigated using a quantum $\mathcal{PT}$-symmetric system assisted by weak measurement, where the authors reported an enhancement of 8.856 times over the conventional Hermitian sensor. However, it should be noted that this enhancement is largely due to the fact that the sensing scheme for the Hermitian sensor was not optimized. Additionally, non-Hermitian lattice systems have been proposed as potential platforms for realizing sensing schemes utilizing the skin effect \cite{budich2020non, koch2022quantum} or the non-reciprocity \cite{mcdonald2020exponentially}, which claimed a sensitivity with an exponential scaling to the chain size. According to our theory, there should be no dependence on the chain size for the ultimate sensitivity, as it only depends on the dimension of the subsystem that directly couples to the parameter being measured.

Although our work demonstrates that coupling to the environment cannot improve the ultimate sensitivity, when the probe state or the measurement protocol is restricted, adding appropriate auxiliary Hamiltonian may be helpful for approaching the ultimate sensitivity bound~\cite{de2013quantum,mishra2021driving,ding2022dynamic}.
In fact, when the parameter couples to the environment, the bounds presented in Eq.~(\ref{eq:estimation_bound}) and~(\ref{eq:qfi_rate}) remain applicable, albeit $\partial H_\lambda/\partial \lambda$ now depends on the environment’s degrees of freedom.
%Besides, although our study focuses on non-Hermitian quantum sensors, it is interesting to scrutinize non-Hermitian sensors based on classical or quasiclassical systems ~\cite{review2020jan} from the perspective of the invariance of information.
In addition, while our study focuses on non-Hermitian sensors implemented by full quantum systems~\cite{xiao2019observation,naghiloo2019quantum,huang2019simulating}, scrutinizing non-Hermitian sensors based on classical or quasiclassical systems~\cite{review2020jan} through the perspective of conservation of information is a compelling avenue for future research.

%Our work gives the sensitivity bound for non-Hermitian sensors implemented by using quantum systems. Although the Hamiltonian extension cannot improve the ultimate sensitivity, when the probe state or the measurement protocol is restricted, adding appropriate auxiliary Hamiltonian may be helpful for approaching the ultimate sensitivity bound~\cite{de2013quantum,ding2022dynamic}.
%Thus it is interesting to scrutinize the non-Hermitian sensors based on classical or quasiclassical systems ~\cite{review2020jan} from the perspective of quantum information.

\begin{acknowledgments}
The work is supported by National Key Research and Development Program of China (Grant No. 2021YFA1402104), the  NSFC under Grants No.12174436, No.11935012
and No.T2121001 and the Strategic Priority Research Program of Chinese Academy of Sciences under Grant No. XDB33000000.
\end{acknowledgments}

\end{document}

\preprint{APS/123-QED}

\title{\textit{Supplementary Material:} Fundamental Sensitivity Limits for non-Hermitian Quantum Sensors}
%\title{Sensitivity bound of non-Hermitian sensors}
%\title{The problem of sensitivity enhancement of non-Hermitian sensors}
%\title{Determination of the coupling strength and the coupling anisotropy in a two-qubit system}% Force line breaks with \\

%\author{Wenkui Ding}
%%\email{wenkuiding@zju.edu.cn}
%\affiliation{Zhejiang Institute of Modern Physics and Department of Physics,
%Zhejiang University, Hangzhou, Zhejiang 310027, China}

%\author{Wenkui Ding}
%%\email{wenkuiding@zju.edu.cn}
%\affiliation{Zhejiang Institute of Modern Physics and Department of Physics,
%Zhejiang University, Hangzhou, Zhejiang 310027, China}
\author{Wenkui Ding}
\affiliation{Beijing National Laboratory for Condensed Matter Physics, Institute of
Physics, Chinese Academy of Sciences, Beijing 100190, China}
\affiliation{Department of Physics, Zhejiang Sci-Tech University, 310018 Zhejiang, China}
\author{Xiaoguang Wang}
\email{xgwang@zstu.edu.cn}
\affiliation{Department of Physics, Zhejiang Sci-Tech University, 310018 Zhejiang, China}
\author{Shu Chen}
\email{schen@iphy.ac.cn}
\affiliation{Beijing National Laboratory for Condensed Matter Physics, Institute of
Physics, Chinese Academy of Sciences, Beijing 100190, China}
\affiliation{School of Physical Sciences, University of Chinese Academy of Sciences,
Beijing, 100049, China}

%\begin{abstract}
%In this paper, we determine the fundamental limits for the sensitivity of the non-Hermitian sensors from the perspective of quantum information.
%We prove that, the non-Hermitian sensors will not outperform their Hermitian counterparts (directly couples to the parameter) in the performance of sensitivity, due to the invariance of the quantum information about the parameter.
%We investigate various non-Hermitian sensing proposals, which are implemented using full quantum systems, and we find that the sensitivity of these sensors is in agreement with our predictions.
%We provide a deeper theoretical understanding of the non-Hermitian enhanced sensing in the quantum domain and build the bridge over the gap between non-Hermitian physics and quantum metrology.
%\end{abstract}

\date{\today}% It is always \today, today,
\maketitle

\section{the bound for the change rate of the quantum Fisher information}
Firstly, we consider the situation that the parameter to be estimated is a multiplicative factor in the Hamiltonian.
The probe state is $|\Psi_0\rangle$ and the Hamiltonian to encode the parameter is $H_\lambda=\lambda H_1+H_0$.
After the time evolution, $|\Psi_f\rangle=e^{-iH_\lambda t}|\Psi_0\rangle$, the final state is dependent on the parameter $\lambda$.
The quantum Fisher information for a pure quantum state is,
\begin{equation}
F_\lambda(t)=4(\langle \Psi_f|\overleftarrow{\frac{\partial}{\partial\lambda}}\overrightarrow{\frac{\partial}{\partial\lambda}}|\Psi_f\rangle-|\langle \Psi_f|\overrightarrow{\frac{\partial}{\partial\lambda}}|\Psi_f\rangle|^2).
\end{equation}
We can also calculate the quantum Fisher information in terms of the initial state as,
\begin{equation}
F_\lambda(t)=4(\langle \Psi_0|h_\lambda^2|\Psi_0\rangle-|\langle \Psi_0|h_\lambda|\Psi_0\rangle|^2)\equiv 4\left.\text{Var}[h_\lambda]\right|_{|\Psi_0\rangle},
\end{equation}
where we have defined the variance of operators and the transformed local generator is defined as,
\begin{equation}
h_\lambda\equiv ie^{iH_\lambda t}\frac{\partial}{\partial \lambda}e^{-iH_\lambda t}=iU_\lambda^\dagger\frac{\partial}{\partial\lambda}U_\lambda.
\end{equation}
According to Ref.~\cite{generalized2007boixo}, we can prove that,
\begin{equation}
F_\lambda^{1/2}\leq t||H_1||,
\end{equation}
where the seminorm $||H_1||\equiv u_{\text{max}}-u_{\text{min}}$, with $u_{\text{max}}$ and $u_{\text{min}}$ the largest and smallest eigenvalue of $H_1$, respectively.
Now, we want to prove that,
\begin{equation}
\left|\frac{\partial F_\lambda^{1/2}}{\partial t}\right|\leq ||H_1||,
\end{equation}
namely the change rate of the quantum Fisher information (or the magnitude of the quantum Fisher information flow) is bounded by the spectral width of the Hamiltonian that directly couples to the parameter.
Calculating the derivative $\left|\frac{\partial F_\lambda^{1/2}}{\partial t}\right|=\frac{1}{2}F_\lambda^{-1/2}\left|\frac{\partial F_\lambda}{\partial t}\right|$, and using the fact that $F_\lambda$ is always a positive quantity, equivalently, we are about to prove that,
\begin{equation}
\left|\frac{\partial F_\lambda}{\partial t}\right|\leq 2F_\lambda^{1/2}||H_1||.
\end{equation}

\textit{Proof}. we have the Schr\"{o}dinger equation as,
\begin{equation}
i\frac{\partial U_\lambda}{\partial t}=H_\lambda U_\lambda, 
\end{equation}
namely, we have the relation that $\frac{\partial U_\lambda}{\partial t}=-iH_\lambda U_\lambda$ and $\frac{\partial U_\lambda^\dagger}{\partial t}=iU_\lambda^\dagger H_\lambda$.
Then, according to the definition of the transformed local generator, we have,
\begin{equation}
\begin{aligned}
\frac{\partial h_\lambda}{\partial t}&=i\frac{\partial U_\lambda^\dagger}{\partial t}\frac{\partial}{\partial\lambda}U_\lambda+iU_\lambda^\dagger\frac{\partial}{\partial\lambda}\frac{\partial U_\lambda}{\partial t}\\
&=i(i U_\lambda^\dagger H_\lambda)\frac{\partial}{\partial\lambda}U_\lambda+iU_\lambda^\dagger\frac{\partial}{\partial\lambda}(-iH_\lambda U_\lambda)\\
&=-U_\lambda^\dagger H_\lambda\frac{\partial U_\lambda}{\partial\lambda}+U_\lambda^\dagger\frac{\partial H_\lambda}{\partial\lambda}U_\lambda+U_\lambda^\dagger H_\lambda\frac{\partial U_\lambda}{\partial\lambda}\\
&=U_\lambda^\dagger\frac{\partial H_\lambda}{\partial\lambda}U_\lambda.
\end{aligned}
\end{equation}
Here, for $H_{\lambda}=\lambda H_1+H_0$, we have $\frac{\partial h_\lambda}{\partial t}=U_\lambda^\dagger H_1 U_\lambda$.

According to the definition of the quantum Fisher information in the form of the variance of the transformed local generator, we obtain,
\begin{equation}
\begin{aligned}
&\frac{\partial F_\lambda}{\partial t}\\
&=4(\langle \Psi_0|\frac{\partial h_\lambda}{\partial t}h_\lambda+h_\lambda\frac{\partial h_\lambda}{\partial t}|\Psi_0\rangle-2\langle \Psi_0|\frac{\partial h_\lambda}{\partial t}|\Psi_0\rangle\langle \Psi_0|h_\lambda|\Psi_0\rangle)\\
&\equiv 8\left.\text{Cov}[\frac{\partial h_\lambda}{\partial t},h_\lambda]\right|_{|\Psi_0\rangle},
\end{aligned}
\end{equation}
where we have defined the covariance of operators as follows,
\begin{equation}
\left.\text{Cov}[\hat{A},\hat{B}]\right|_{|\Psi\rangle}=\frac{1}{2}\langle \Psi|\hat{A} \hat{B}+\hat{B} \hat{A} |\Psi\rangle-\langle \Psi|\hat{A} |\Psi\rangle\langle \Psi|\hat{B} |\Psi\rangle.
\end{equation}
Using Cauchy-Schwarz inequality, we can prove the following relation (for compactness we neglect the subscript $|\Psi\rangle$ in the following discussions),
%\begin{equation}
%\left|\left.\text{Cov}(\hat{A},\hat{B})\right|_{|\Psi\rangle}\right|\leq \sqrt{\left.\text{Var}(\hat{A})\right|_{|\Psi\rangle}\left.\text{Var}(\hat{B})\right|_{|\Psi\rangle}}.
%\end{equation}
\begin{equation}
\label{eq:covariance}
\left|\text{Cov}[\hat{A},\hat{B}]\right|\leq \sqrt{\text{Var}(\hat{A})\text{Var}(\hat{B})}.
\end{equation}
To prove this inequality, we define,
\begin{equation}
\hat{C}=\hat{A}-\frac{\text{Cov}[\hat{A},\hat{B}]}{\text{Var}[\hat{B}]}\hat{B}.
\end{equation}
Then, we have,
\begin{equation}
\begin{aligned}
&\text{Var}[\hat{C}]=\langle \Psi|\hat{A}^2|\Psi\rangle+\langle \Psi|\hat{B}^2|\Psi\rangle\left(\frac{\text{Cov}[\hat{A},\hat{B}]}{\text{Var}[\hat{B}]}\right)^2\\
&\text{ }-\langle \Psi|\hat{A}\hat{B}+\hat{B}\hat{A}|\Psi\rangle\frac{\text{Cov}[\hat{A},\hat{B}]}{\text{Var}[\hat{B}]}-\langle \Psi|\hat{B}|\Psi\rangle^2\left(\frac{\text{Cov}[\hat{A},\hat{B}]}{\text{Var}[\hat{B}]}\right)^2\\
&\text{ }-\langle \Psi|\hat{A}|\Psi\rangle^2+2\langle \Psi|\hat{A}|\Psi\rangle\langle \Psi|\hat{B}|\Psi\rangle\frac{\text{Cov}[\hat{A},\hat{B}]}{\text{Var}[\hat{B}]}\\
&=\text{Var}[\hat{A}]+\text{Var}[\hat{B}]\left(\frac{\text{Cov}[\hat{A},\hat{B}]}{\text{Var}[\hat{B}]}\right)^2-2\text{Cov}[\hat{A},\hat{B}]\frac{\text{Cov}[\hat{A},\hat{B}]}{\text{Var}[\hat{B}]}\\
&=\text{Var}[\hat{A}]-\frac{(\text{Cov}[\hat{A},\hat{B}])^2}{\text{Var}[\hat{B}]}.
\end{aligned}
\end{equation}
Since $\text{Var}[\hat{C}]\geq 0$, the inequality in Eq.~\ref{eq:covariance} is proved.
As a result, we have that,
\begin{equation}
\begin{aligned}
\left|\text{Cov}[\frac{\partial h_\lambda}{\partial t},h_\lambda(t)]|_{|\Psi_0\rangle}\right|&\leq \sqrt{\text{Var}[\frac{\partial h_\lambda}{\partial t}]|_{|\Psi_0\rangle}\text{Var}[h_\lambda(t)]|_{|\Psi_0\rangle}}\\
&= \sqrt{\text{Var}[U_\lambda^\dagger H_1 U_\lambda]|_{|\Psi_0\rangle}}\frac{F_\lambda^{1/2}(t)}{2}\\
&\leq\frac{||H_1||}{2}\frac{F_\lambda^{1/2}(t)}{2},
\end{aligned}
\end{equation}
where we have used the relation that $F_\lambda(t)=4\left.\text{Var}[h_\lambda(t)]\right|_{|\Psi_0\rangle}$ and $\text{Var}[U_\lambda^\dagger H_1 U_\lambda]\leq\frac{||U_\lambda^\dagger H_1 U_\lambda||^2}{4}=\frac{||H_1||^2}{4}$, namely the variance of the Hermitian operator is bounded by its spectral width and the unitary evolution will not change the spectrum.
Finally, we obtain,
\begin{equation}
\left|\frac{\partial F_\lambda}{\partial t}\right|\leq 2F_\lambda^{1/2}||H_1||.
\end{equation}

The proof of the general case that the parameter is not a multiplicative factor as shown in the main text is very similar and we now briefly discuss the proof method.
The Hamiltonian to encode the parameter now becomes $H_\lambda(t)=H_1(\lambda,t)+H_0(t)$ and the corresponding evolution operator becomes $U_\lambda=\mathcal{T}e^{-i\int_0^tH_\lambda(s)ds}$.
The transformed local generator is defined as $h_\lambda=iU_\lambda^\dagger\frac{\partial}{\partial\lambda}U_\lambda$.
We already have the bound of the quantum Fisher information as,
\begin{equation}
F_\lambda^{1/2}(t)\leq\int_0^t{\left|\left|\frac{\partial H_\lambda(s)}{\partial \lambda}\right|\right|ds},
\end{equation}
and we want to prove that,
\begin{equation}
\left|\frac{\partial F_\lambda^{1/2}(t)}{\partial t}\right|\leq \left|\left|\frac{\partial H_\lambda(t)}{\partial \lambda}\right|\right|.
\end{equation}
To this end, we just need to follow the same procedure as above, except replacing $H_1$ by $\frac{\partial H_\lambda}{\partial \lambda}$.

\section{Rate of dynamic quantum Fisher information for the pseudo-Hermitian quantum sensor}
For the pseudo-Hermitian quantum sensor, the Hamiltonian to encode the parameter in the enlarged two-qubit system is as follows,
\begin{equation}
\hat{H}_\text{tot}=b\hat{I}^{(a)}\otimes \hat{\sigma}_x^{(s)}-c\hat{\sigma}_y^{(a)}\otimes\hat{\sigma}_y^{(s)}+\lambda\hat{I}^{(a)}\otimes\hat{\sigma}_x^{(s)}.
\end{equation}
The probe state or the initial state is
\begin{equation}
\label{eq:probe_state}
|\Psi_0\rangle=\left(\sqrt{\frac{1+\varepsilon}{1+2\varepsilon}}|0\rangle_a+\sqrt{\frac{\varepsilon}{1+2\varepsilon}}|1\rangle_a\right)\otimes|0\rangle_s.
\end{equation}
The dynamic quantum Fisher information can be calculated using
\begin{equation}
F_\lambda(t)=4(\langle \Psi_0|h_\lambda^2(t)|\Psi_0\rangle-|\langle\Psi_0|h_\lambda(t)|\Psi_0\rangle|^2),
\end{equation}
where the Hermitian operator 
\begin{equation}
h_\lambda(t)\equiv iU_\lambda^\dagger(0\rightarrow t)\frac{\partial}{\partial \lambda}U_\lambda(0\rightarrow t)
\end{equation}
is the transformed local generator.
Since $[\hat{H}_{\text{tot}},\hat{\sigma}_y^{(a)}]=0$, we can represent the Hamiltonian in the two decoupled subspace as follows,
\begin{equation}
\hat{H}_\text{tot}=\left|\uparrow_y\right\rangle\left\langle\uparrow_y\right|\otimes\hat{H}_1+\left|\downarrow_y\right\rangle\left\langle\downarrow_y\right|\otimes\hat{H}_2,
\end{equation}
with,
\begin{equation}
\begin{aligned}
\hat{H}_1&=(b+\lambda)\hat{\sigma}_x^{(s)}-c\hat{\sigma}_y^{(s)},\\
\hat{H}_2&=(b+\lambda)\hat{\sigma}_x^{(s)}+c\hat{\sigma}_y^{(s)}.
\end{aligned}
\end{equation}
As a result, we can prove that,
\begin{equation}
\begin{aligned}
h_\lambda(t)&=ie^{i\hat{H}_\text{tot}t}\frac{\partial}{\partial\lambda}e^{-i\hat{H}_\text{tot}t}\\
&=\left|\uparrow_y\right\rangle\left\langle\uparrow_y\right|\otimes \hat{h}_1+\left|\downarrow_y\right\rangle\left\langle\downarrow_y\right|\otimes \hat{h}_2,
\end{aligned}
\end{equation}
where the transformed local generator in the respective subspace is as follows,
\begin{equation}
\begin{aligned}
\hat{h}_1&\equiv ie^{i\hat{H}_1t}\frac{\partial}{\partial\lambda}e^{-i\hat{H}_1t}=h_x\hat{\sigma}_x^{(s)}+h_y\hat{\sigma}_y^{(s)}+h_z\hat{\sigma}_z^{(s)},\\
\hat{h}_2&\equiv ie^{i\hat{H}_2t}\frac{\partial}{\partial\lambda}e^{-i\hat{H}_2t}=h_x\hat{\sigma}_x^{(s)}-h_y\hat{\sigma}_y^{(s)}-h_z\hat{\sigma}_z^{(s)},
\end{aligned}
\end{equation}
with the coefficients,
\begin{equation}
\begin{aligned}
h_x&=\frac{t(\lambda+b)^2}{\Omega^2}+\frac{c^2\sin(2\Omega t)}{2\Omega^3},\\
h_y&=\frac{c(\lambda+b)}{\Omega^2}\left[\frac{\sin(2\Omega t)}{2\Omega}-t\right],\\
h_z&=-\frac{c\sin^2(\Omega t)}{\Omega^2},
\end{aligned}
\end{equation}
with $\Omega=\sqrt{(\lambda+b)^2+c^2}$.
Meanwhile, we can represent the initial state in the eigenbasis of $\hat{\sigma}_y^{(a)}$ as follows,
\begin{equation}
|\Psi_0\rangle=\left(\alpha\left|\uparrow_y\right\rangle+\beta\left|\downarrow_y\right\rangle\right)\otimes|0\rangle_s.
\end{equation}
Then we have,
\begin{equation}
\begin{aligned}
F_\lambda(t)=&4\left(\langle \Psi_0|h_\lambda^2(t)|\Psi_0\rangle-|\langle\Psi_0|h_\lambda(t)|\Psi_0\rangle|^2\right)\\
=&4\left(\alpha^2{}_s\langle 0|\hat{h}_1^2|0\rangle_s+\beta^2{}_s\langle 0|\hat{h}_2^2|0\rangle_s\right)\\
&-4\left(\alpha^2{}_s\langle 0|\hat{h}_1|0\rangle_s+\beta^2{}_s\langle 0|\hat{h}_2|0\rangle_s\right)^2.
\end{aligned}
\end{equation}
Using the above equations, we have
\begin{equation}
\begin{aligned}
{}_s\langle 0|\hat{h}_1|0\rangle_s&=h_z,\\
{}_s\langle 0|\hat{h}_2|0\rangle_s&=-h_z,\\
{}_s\langle 0|\hat{h}_1^2|0\rangle_s&=h_x^2+h_y^2+h_z^2,\\
{}_s\langle 0|\hat{h}_2^2|0\rangle_s&=h_x^2+h_y^2+h_z^2.
\end{aligned}
\end{equation}
Therefore, the dynamic quantum Fisher information is,
\begin{equation}
F_\lambda(t)=4\left[h_x^2+h_y^2+h_z^2-(\alpha^2-\beta^2)h_z^2\right].
\end{equation}
For the probe state $|\Psi_0\rangle$ in Eq.~(\ref{eq:probe_state}), we can check $\alpha^2-\beta^2=0$, thus the dynamic quantum Fisher information becomes
\begin{equation}
F_\lambda(t)=\frac{4(b+\lambda)^2}{\Omega^2}t^2+\frac{4c^2\sin^2(\Omega t)}{\Omega^4}.
\end{equation}
Finally, the change rate of the dynamic quantum Fisher information can be exactly calculated as follows,
\begin{equation}
\label{eq:rate_QFI}
\frac{\partial F^{1/2}_\lambda(t)}{\partial t}=2\frac{\cos^2\theta+\sin^2\theta\frac{\sin{(2\Omega t)}}{2\Omega t}}{\sqrt{\cos^2\theta+\sin^2\theta\frac{\sin^2(\Omega t)}{(\Omega t)^2}}},
\end{equation}
where we have defined $(b+\lambda)/\Omega=\cos\theta$ and $c/\Omega=\sin\theta$.
From Eq.~(\ref{eq:rate_QFI}), it follows that $-2\leq\frac{\partial F^{1/2}_\lambda(t)}{\partial t}\leq 2$, which verifies our theory in the main text.

\section{the population fluctuation for the two-level system}
We now briefly discuss the quantum projection noise in the population measurement of the two-level system~\cite{itano1993quantum}.
The quantum state that needs to be measured is,
\begin{equation}
|\psi\rangle=\alpha\left|\uparrow\right\rangle+\beta\left|\downarrow\right\rangle.
\end{equation}
For this state, we have the population (probability) in the $\left|\uparrow\right\rangle$ state after the projection measurement as, $p_\uparrow=|\alpha|^2$.
After a repetition of $N$ times, the number of measurement outcomes to be the $\left|\uparrow\right\rangle$ state is $N_\uparrow$, then we can estimate the probability $p_\uparrow\approx p_\uparrow^\prime=N_\uparrow/N$.
Since $N$ is finite, there are uncertainties in the value of $N_\uparrow$ and equivalently there are uncertainties in the determination of $p_\uparrow^\prime$.
Since the repeated projection process follows the binomial distribution, the variance of the times that projected to the up state is $\text{Var}(N_\uparrow)=Np_\uparrow(1-p_\uparrow)$. 
Correspondingly, the variance of the estimated population is,
\begin{equation}
\text{Var}(p_\uparrow^\prime)=\text{Var}(\frac{N_\uparrow}{N})=\frac{p_\uparrow(1-p_\uparrow)}{N}.
\end{equation}

The discussion above can be extended to the projective measurement of multi-level quantum systems.
As an example, for a general two-qubit quantum state,
\begin{equation}
|\psi\rangle=\alpha_1|00\rangle+\alpha_2|01\rangle+\alpha_3|10\rangle+\alpha_4|11\rangle,
\end{equation}
the distribution of the times $N_1$, corresponding to the projected outcome in $|00\rangle$, for $N$ repeated measurements follows the multinomial distribution.
Thus, the variance of $N_1$ is,
\begin{equation}
\text{Var}(N_1)=Np_1(1-p_1),
\end{equation}
where $p_1=|\alpha_1|^2$.
Accordingly, the uncertainty of the estimated population is,
\begin{equation}
\text{Var}(p_1^\prime)=\text{Var}(\frac{N_1}{N})=\frac{p_1(1-p_1)}{N}.
\end{equation}